\documentclass[12pt]{article}
\usepackage{epsfig,latexsym}

\input epsf

\begin{document}

\font\cmss=cmss10 \font\cmsss=cmss10 at 7pt 
\hfill NYU-TH/00/01/02

\hfill CERN-TH/2000-041

\hfill Bicocca-FT/00/04 \vskip .1in \hfill hep-th/0002066

\hfill

\vspace{20pt}

\begin{center}
{\Large \textbf{A NOTE ON THE HOLOGRAPHIC BETA AND $C$ FUNCTIONS}}
\end{center}

\vspace{6pt}

\begin{center}
\textsl{D. Anselmi $^a$, L. Girardello $^{b,d}$, M. Porrati $^c$ and A.
Zaffaroni $^{d}$} \vspace{20pt}

$^{a}$\textit{CERN, Division Th\'{e}orique, CH-1211, Geneva 23, Switzerland }

\textit{$^{b}$ Universit\`{a} di Milano-Bicocca, Dipartimento di Fisica}

\textit{$^c$ Department of Physics, NYU, 4 Washington Pl, New York NY 10012}

\textit{$^{d}$INFN - Sezione di Milano, Via Celoria 16, Milan, Italy}
\end{center}

\vspace{12pt}

\begin{center}
\textbf{Abstract }
\end{center}

\vspace{4pt} {\small \noindent The holographic RG flow in AdS/CFT
correspondence naturally defines a holographic scheme in which the central
charge $c$ and the beta function are related by the formula $\dot{c}%
=-2c\beta_a\beta_bG^{ab}$, where $G^{ab}$ is the metric of the kinetic term
of the supergravity scalars. In particular, the metric in the space of
couplings is $f^{ab}=2cG^{ab}$. We perform some checks of that result and we
compare it with the quantum field theory expectations. We discuss
alternative definitions of the $c$-function. In particular, we compare, for
a particular supersymmetric flow, the holographic $c$-function with the
central charge computed directly from the two-point function of the
stress-energy tensor.}

\vfill\eject 
\noindent

Conformal field theories in four dimensions have two main central charges, $c
$ and $a$, which multiply the square of the Weyl tensor and the Euler
density, respectively, in the trace anomaly. In a quantum field theory
interpolating between UV and IR conformal fixed points the total flows of $c$
and $a$, i.e. the differences $\Delta c=c_{\mathrm{UV}}-c_{\mathrm{IR}}$ and 
$\Delta a=a_{\mathrm{UV}}-a_{\mathrm{IR}}$, give important physical
information (see for instance \cite{afgj}). The flows can be induced by
dimensionful parameters, by
the renormalization-group scale $\mu $, or by the combined effect of both. 

The RG flow, induced by $\mu $, is irreversible, which means that it
satisfies the inequality $\Delta a\geq 0$. The irreversibility of the RG
flow is better studied when dimensionful parameters are absent. This means
that the theory is conformal at the classical level. In \cite{a2} a
non-perturbative formula for the RG flow $\Delta a$ was obtained and checked
in perturbation theory.

The flows induced by relevant deformations have a
quantitatively different effect on $\Delta a,$ although they still obey the
inequality $\Delta a\geq 0$. Nevertheless, there is a special class of
theories in every even dimension, the theories interpolating
between $c=a$ fixed points, where the formula for $%
\Delta a$ (equal to $\Delta c$) is universal \cite{a1}. This universality
also holds in two dimensions. The even-dimensional conformal field theories
with $c=a$ share various properties with two dimensional conformal field
theory \cite{a1}.

The ``holographic'' supergravity/gauge theory correspondence considers, in
the 5-d gauged supergravity limit, precisely a class of $c=a$ conformal
field theories \cite{hs}, the simplest example being the N=4
supersymmetric Yang-Mills theory in the strongly coupled large-$N$
limit. Other examples have been
constructed in the literature and need not be supersymmetric.

In ~\cite{gppz1,gppz2,gppz3,dz,freed1,freed2} flows induced by massive
deformations were considered in the context of this correspondence. We call
them the ``holographic'' flows. On the basis of the considerations recalled
above, we expect, and are indeed going to check in the present paper, that:

1) The ``holographic'' $c$-function defined in \cite{gppz1} obeys a formula
similar to the formula for the RG\ flow of the $a$-function in quantum field
theory. This is our result (\ref{noncanonica}). In particular, the
holographic central charge $c$ is stationary at the fixed points. Observe
that the stationarity of $c$ is not true in a general quantum field theory
(with $c\neq a$ at the fixed points) and is peculiar
of the holographic flows;

2) Other definitions of $c$, not related to the equality $c=a$, but equally
convenient in quantum field theory, for example the central function defined
by the stress-tensor two-point function, should exhibit similar properties:
monotonicity and stationarity at the fixed points. These facts are also
peculiar of the holographic flows, because it is well known that $c$ does
not even decrease in a general quantum field theory.

\medskip

We begin by discussing the properties of the $c$-function proposed in ref.~%
\cite{gppz1} and work out the general formula for its derivative $\dot c$
along the flow. Secondly, we directly compute the $c$-function using the
correlator of two stress-energy tensors, always using the holographic
correspondence, and compare the two definitions. We explain why the two
definitions are compatible (in particular, both positive and interpolating
monotonically between the critical values) even though they are not equal.

A candidate $c$-function, decreasing along the holographic flow was proposed
in~\cite{gppz1,freed1}. In the notation of \cite{gppz1} the $c$-function is: 
\begin{equation}
c=\mathrm{const.}\, (T_{yy})^{-3/2} =\left( \frac{\mathrm{d}\phi }{\mathrm{d}%
y}\right) ^{-3},  \label{cfunz}
\end{equation}
Where $\phi$ is the scale factor of the 5-d supergravity metric, and $y$ is
its radial coordinate: $ds^2=dy^2+\exp(2\phi)dx_\mu dx^\mu$.

The equations for a holographic RG flow generated by one of the
perturbations that can be studied within 5-d gauged supergravity are, in the
notations of ref.~\cite{gppz1}~\footnotemark : 
\begin{equation}
\frac{\mathrm{D}}{\mathrm{D}y}\left(e^{4\phi} G_{ab}\frac{\mathrm{d}\lambda^b%
}{\mathrm{d}y} \right) =e^{4\phi}{\frac{\partial V}{\partial\lambda^a}}%
,\qquad 6\left( \frac{\mathrm{d} \phi}{\mathrm{d}y }\right) ^{2}=\sum_{ab}
G_{ab}\frac{\mathrm{d}\lambda^a}{\mathrm{d}y } \frac{\mathrm{d}\lambda^b}{%
\mathrm{d}y }-2V.  \label{holorg}
\end{equation}
Here $\lambda^a$ denotes the 42 scalars of 5-d N=8 gauged supergravity, $%
G_{ab}$ denotes the metric of their kinetic term and D/D$y$ is the covariant
derivative. From now on we will set for simplicity and with no loss of
generality $G_{ab}=\delta_{ab}$. 
\footnotetext{
These perturbations have UV dimension 2 or 3; in gauge theory, they
correspond to mass terms for scalars and/or fermions, and trilinear terms in
the scalar potential. In supergravity, they correspond to VEVs of some of
the 42 scalars in the 5-d, N=8 supergravity multiplet.}

These equations imply, in particular, that the second derivative of $\phi$
does not depend on the potential $V$: 
\[
\frac{\mathrm{d}^{2}\phi }{\mathrm{d}y^{2}}=-\frac{2}{3}\sum_{a}\left( \frac{%
\mathrm{d}\lambda _{a}}{\mathrm{d}y}\right) ^{2}. 
\]
We also have 
\[
\frac{\mathrm{d}c}{\mathrm{d}\phi }=-3\left( \frac{\mathrm{d}\phi }{\mathrm{d%
}y}\right) ^{-5}\frac{\mathrm{d}^{2}\phi }{\mathrm{d}y^{2}}=2c\sum_{a}\left( 
\frac{\mathrm{d}\lambda _{a}}{\mathrm{d}\phi }\right) ^{2}. 
\]
To obtain quantitative agreement with QFT results (see \cite{a2}) and a
consistent picture of the holographic RG flow we must set 
\begin{equation}
\phi =\ln \mu ,\qquad \beta _{a}=\frac{\mathrm{d}\lambda _{a}}{\mathrm{d}%
\phi }.  \label{beta}
\end{equation}
Therefore: 
\begin{equation}
\dot{c}=-{\frac{\mathrm{d}c}{\mathrm{d}\phi}}=-2c\sum_{a}\beta _{a}^{2}.
\label{roa}
\end{equation}

Let us recall a few other results from quantum field theory \cite{agj}.
Defining 
\begin{equation}
\Theta =\beta _{a}\mathcal{O}_{a}  \label{uno}
\end{equation}
and 
\[
\mu {\frac{\mathrm{d}}{\mathrm{d\mu }}}\beta _{a}=-\dot{\beta}_{a}=\Delta
_{ab}\beta _{b}, 
\]
a theorem proved in \cite{agj} states that the critical value $h_{*}$ of the 
$\Theta $-anomalous dimension, 
\[
\langle \Theta (x)~\Theta (0)\rangle =\frac{\mathrm{const.}}{|x|^{8+2h_{*}}}%
, 
\]
equals the minimal real part of the $\Delta$-eigenvalues, in the IR limit,
and the maximal real part of the $\Delta$-eigenvalues in the UV limit. Note
that $h_{*}$ is also the anomalous dimension of the off-critical deformation
of the theory, i.e. the operator $\lambda _{a}\mathcal{O}_{a}$ (the
deformation being $\mathcal{L}_{*}\rightarrow \mathcal{L=L}_{*}+\lambda _{a}%
\mathcal{O}_{a}$, where $\mathcal{L}_{*} $ denotes the critical Lagrangian).

\bigskip

Formula (\ref{roa}) implies, in particular, 
\begin{equation}
-\frac{\ddot{c}}{2\dot{c}}=\frac{\sum_{a,b}\beta _{a}\Delta _{ab}\beta _{b}}{
\sum_{a}\beta _{a}^{2}}+\sum_{a}\beta _{a}^{2}.  \label{limit}
\end{equation}
At criticality the second term vanishes, while the first term selects the
minimal- or maximal-real-part eigenvalue of the matrix $\Delta _{ab}$, as we
now show. Note that $\Delta _{ab}$ is in general not symmetric. We can
diagonalize it in a complex space. Let $\Delta=P^{-1}DP$, with $D=\mathrm{%
diag}(\delta_a)$, $\delta _{a}$ denoting the eigenvalues. Let us write,
around the critical point, 
\[
\beta _{a}(\lambda )=\Delta _{ab}\lambda _{b},\qquad \beta _{a}(\mu )=\Delta
_{ab}\,\mu ^{\Delta _{bc}}k_{c}=(P^{-1}D\mu^{D}Pk)_{a}, 
\]
$k_{c}$ denoting arbitrary constants. Now, in the UV limit ($\mu \rightarrow
\infty $) the behavior of the first term of (\ref{limit}) is dominated by
the eigenvalue of the matrix $\Delta _{ab}$ with maximal real part. It is
dominated by the eigenvalue with minimal real part in the IR limit ($\mu
\rightarrow 0 $). The imaginary parts of the eigenvalues are irrelevant
phases. In conclusion, we have 
\[
-\frac{\ddot{c}}{2\dot{c}}=\max \mbox{Re\,}\delta _{a}~~~~ \mathrm{%
in\,the\,UV},~\qquad ~-\frac{\ddot{c}}{2\dot{c}}=\min \mbox{Re\,}\delta
_{a}~~~~\mathrm{in\,the\,IR}. 
\]
These are also the values of the anomalous dimensions of the operators $%
\mathcal{O}_{a}$ at criticality, as proved in \cite{agj}. Therefore we have,
in complete generality, 
\begin{equation}
h_{*}=-\lim_{*}\frac{\ddot{c}}{2\dot{c}},  \label{accastar}
\end{equation}
where the star denotes criticality.

The ``anomalous dimension'' $h_*$ denotes the deviation of the total
dimension from the reference value 4, $h_*=\Delta-4$ in the conventional
notation. We can check, in complete generality, that this quantum field
theoretical prediction is correctly reproduced by the holographic flows.
Indeed, the second equation of (\ref{holorg}) implies that, around a fixed
point, 
\[
\frac{\mathrm{d}\phi}{\mathrm{d}y }={\frac{1}{R}}, 
\]
$R$ being the AdS radius, and the first of eqs.~(\ref{holorg}) gives 
\[
\lambda\sim \mathrm{const.}\, \mathrm{e}^{-(4-\Delta)y/R}=\mathrm{const.}\, 
\mathrm{e}^{-(4-\Delta)\phi} 
\]

At this point, it is straightforward to see that (\ref{accastar}) gives $%
\Delta -4$. The same can be see from the definition of $\beta$ in (\ref{beta}%
), confirming that the natural definition of holographic beta function works
correctly.

All the results described above generalize to non-canonical scalar metrics,
in particular 
\begin{equation}
\dot{c}=-2cG^{ab}\beta_a\beta_b.  \label{noncanonica}
\end{equation}

We must remark that our definition of $c$ is unique only at the critical
points $\dot{c}=0$. Away from criticality, $c$ need not coincide with
central functions defined in other ways; indeed, it need not coincide with
other holographic definitions of $c$, as for instance that given in ref.~%
\cite{dbvv}. This non-uniqueness even within the holographic scheme follows
from the ambiguity in the identification of $\phi$ as a function of the
scale $\mu$. Only at the critical points, $\dot{c}=0$, is the standard
identification, $\phi = \log(\mu/\mu_0)$, unique, because of the AdS/CFT
correspondence. Away from criticality, uniqueness is lost.

A canonical definition of $c$ as a function of the scale is obtained by
computing the two point function of the stress-energy tensor using the
equation~\cite{a3} 
\begin{equation}
\langle T_{\mu\nu} (x) T_{\rho\sigma}(0)\rangle = -{\frac{1}{48\pi^4}}{\prod}%
^{(2)}_{\mu\nu\rho\sigma} \left[{\frac{c(x)}{x^4}}\right] +
\pi_{\mu\nu}\pi_{\rho\sigma}\left[{\frac{f(x)}{x^4}}\right],
\label{duepunti}
\end{equation}
where $\pi_{\mu\nu}=\partial_\mu\partial_\nu -\eta_{\mu\nu}\partial^2 $, and 
${\prod}^{(2)}_{\mu\nu\rho\sigma}=2\pi_{\mu\nu}\pi_{\rho\sigma} -3
(\pi_{\mu\rho}\pi_{\nu\sigma}+\pi_{\mu\sigma}\pi_{\nu\rho})$. We will call
this $c$ the \textit{canonical} $c$-function.

For a generic flow, it is impossible to compute analytically this two-point
function, even in the supergravity approximation. To the best of our
knowledge, there are few exceptions, namely, the solutions describing the
Coulomb branch of the N=4 supersymmetric gauge theory~\cite{freed2} and the
N=1 supersymmetric flows studied in ref.~\cite{gppz3}, which interpolate
between the N=4 UV theory and an IR N=1 pure super Yang-Mills theory.

Here, we mostly consider the flow to pure N=1 YM theory. Only a few
modifications of the computation described below are required to study the
N=4 Coulomb branch, which will be briefly discussed at the end of this paper.

The flow we shall consider corresponds to an IR vacuum with zero gaugino
condensate. By rescaling the AdS radius to $R=1$ and setting the IR
singularity of the metric at $y=0$ (see ref.~\cite{gppz3} for details), the
5-d metric is completely specified by the scale factor 
\begin{equation}
e^{2\phi(y)}= e^{2y} -1.  \label{metrica}
\end{equation}
To compute the two-point function of the transverse-traceless part of the
stress-energy tensor using the holographic correspondence, we need to solve
the linearized equations of motion for the 5-d graviton on the background
specified by eq.~(\ref{metrica}). These equations simplify dramatically for
the transverse-traceless part, when they become identical with the equations
of motion of a minimally-coupled massless scalar, denoted here by $\chi(x)$.
By writing $\chi(x)=\exp(ik_\mu x^\mu)\chi_k$ we find 
\begin{equation}
-{\frac{\mathrm{d}^2 }{dy^2}}\chi_k - 4{\frac{d\phi}{dy}} {\frac{\mathrm{d}}{%
dy}}\chi_k + k^2e^{-2\phi}\chi_k=0.  \label{eom}
\end{equation}
With the change of variable $x=\exp(-2y)$, a few elementary algebraic
manipulations, and dropping the label $k$, eq.~(\ref{eom}) reduces to a
standard hypergeometric equation 
\begin{equation}
x(1-x){\frac{\mathrm{d}^2 }{dy^2}}\chi -(1+x) {\frac{\mathrm{d}}{dy}}\chi +
a^2\chi=0,\;\;\; a^2\equiv -{\frac{k^2}{4}},  \label{ipergeo}
\end{equation}
whose two solutions are (cfr. \cite{as} for notations) 
\begin{eqnarray}
\chi_1&=& x^2F(a+2,-a+2;3;x),  \label{sol1} \\
\chi_2&=& x^2\log(x)F(a+2,-a+2;3;x) + \sum_{n=1}^{\infty} [\psi(a+n+2)
-\psi(a+2) +\psi(-a+n+2)+  \nonumber \\
&& -\psi(-a+2) -\psi(n+3) +\psi(3) -\psi(n+1) + \psi(1)]{\frac{%
(a+2)_n(-a+2)_n}{3_n n!}}x^{n+2}+  \nonumber \\
&&+{\frac{4}{a^2(a^2-1)}} -x.  \label{sol2}
\end{eqnarray}
The linear combination of $\chi_1, \chi_2$ regular at $x=1$ and normalized
to $1$ at $x=0$ is 
\begin{equation}
\chi= {\frac{a^2(a^2-1)}{4}}\{\chi_2 - [2-\psi(a+2)-\psi(-a+2) +\psi(3)
+\psi(1)] \chi_1\}.  \label{comblin}
\end{equation}
The two-point function of the stress-energy tensor is extracted from this
expression in the usual manner~\cite{gkp,w}. Namely, we compare eq.~(\ref
{comblin}) with eq.~(\ref{duepunti}), and we normalize the central charge in
the UV using formula~(32) of ref.~\cite{gkp}. To simplify, we choose as in~%
\cite{gkp} a Euclidean 4-momentum ($k^2\geq 0$) oriented along the $z$
coordinate, and we find 
\begin{equation}
\langle \tilde{T}_{xy}(k) T_{xy}(0)\rangle= -{\frac{N^2}{64\pi^2}} k^2
(k^2+4)\mbox{Re\,}\psi(2+ik) + P(k^2).  \label{corifl}
\end{equation}
Here $P(k^2)$ denotes a polynomial in $k^2$ which only contributes to
contact terms. In the UV this formula approaches, obviously, the pure-AdS
form 
\begin{equation}
\langle \tilde{T}_{xy}(k) T_{xy}(0)\rangle= -{\frac{N^2}{64\pi^2}}%
k^4\log(k^2) + \tilde P(k^2).  \label{uvlimit}
\end{equation}
The 2-point function of the transverse-traceless part of $T_{\mu\nu}$ is
proportional to eq.~(\ref{corifl}), as we noticed above. From equation~(\ref
{duepunti}), we can read the $c$-function 
\begin{equation}
\int d^4 xe^{ikx}{\frac{c(x)}{|x|^4}}=-{\frac{\pi^2 N^2}{4}}{\frac{k^2+4}{k^2%
}} \mbox{Re\,}\psi (2+{\frac{ik}{2}})=-{\frac{\pi^2 N^2}{4}}%
\sum_{n=2}^\infty {\frac{k^2+4}{n(4n^2+k^2)}}  \label{conf}
\end{equation}
where we used the series expansion for $\psi$. In the right hand side of
this equation, we discarded any contact or trace term.

The Fourier transform can be inverted (modulo contact terms) to give a
closed expression for $c(x)$ 
\begin{equation}
c(x)={\frac{N^2}{2}}\sum_{n=2}^\infty (n^2-1)|x|^3 K_1(2n|x|)= {\frac{N^2}{2}%
}|x|^3 \int_0^\infty {\frac{3e^{2x\cosh t}-1}{(e^{2x\cosh t}-1)^3}}\cosh t dt
\label{int}
\end{equation}
Every candidate $c$-function has to satisfy some crucial requirements. First
of all, it must be positive definite; this is manifest from equation~(\ref
{int}). Second, it must coincide with the value of the central charge at the
fixed points of the RG group. This is indeed the case. For small $x$, $%
c(x)\rightarrow c_{UV}={N^2/8}$, while for large $x$, $c(x)\sim
x^{5/2}e^{-4x}\rightarrow c_{IR}=0$, as appropriate for a confining theory.
Finally, it must be monotonic. This can be checked by an explicit
computation: 
\begin{equation}
\dot c=x{\frac{dc}{dx}}={\frac{3N^2x^3}{2}}\int_0^\infty {\frac{(3e^{2x\cosh
t}-1) (e^{2x\cosh t}-1)-4x\cosh t e^{4x\cosh t}}{(e^{2x\cosh t}-1)^4}}\cosh
t dt  \label{deriv}
\end{equation}
One can easily check that the integrand is negative definite. This is a
non-trivial result that confirms the interpretation of supergravity
solutions as description of quantum field theory RG flows.

We can compute the first terms in the small $x$-expansion of $c(x)$ 
\begin{equation}
c(x)=N^2\left[{\frac{1}{8}}+{\frac{x^2}{4}}\log x+O(x^2)\right]
\label{expan}
\end{equation}
Inserting this expansion into formula (\ref{accastar}) we find correctly $h_{%
\mathrm{UV}}=-1$, since the deformation is generated by a fermionic mass
term ($\Delta=3$).

The holographic $c$-function for the flow to pure N=1 YM is easily computed
from equation~(\ref{metrica}). With the naive identification $%
\phi=\log(\mu/\mu_0)$, $\mu_0=\mbox{constant}$, one finds 
\begin{equation}
c_{\mathrm{H}}(\mu)={\frac{N^2}{8}}{\frac{\mu^6}{(\mu^2+\mu_0^2)^3}}
\label{chol}
\end{equation}

The $c$-function given by eq.~(\ref{duepunti}), instead, depends
non-analytically on $\mu\propto 1/x$ already at small $x$, as shown by eqs.~(%
\ref{expan}) and~(\ref{chol}). This result does not mean that the two
definitions are incompatible, rather, as pointed out before, it means that
the identification of $\exp(\phi)$ with the scale $\mu$ does not hold
outside the critical points.

The computation of the function $f(x)$ in eq.~(\ref{duepunti}) would be
interesting because, as noticed in ref.~\cite{a2}, it is related to the
derivative of $c(x)$. Unfortunately, the computation of $f(x)$ can not be
reduced to the one for a minimally coupled scalar field and requires the
full stress-energy tensor two point function.

We conclude with a few observations about the holographic scheme.

\textit{i}) An ansatz such as $\dot{c}\sim \beta _{{}}^{k}$ with $k\neq 2$
would disagree with the AdS/CFT correspondence (it would not verify the
check above). It would disagree also with quantum field theory \cite{a2}. In
this sense we have a consistent check of holography versus quantum field
theory.

\textit{ii}) The metric $f$ in the space of couplings \cite{a2} -- i.e. the
higher dimensional analogue of the Zamolodchikov metric -- is the metric of
the 5-d supergravity scalars times $c$ itself: 
\begin{equation}
f_{ab}=2cG _{ab}.  \label{holometric}
\end{equation}
Therefore, the monotonicity of $c$ is directly implied by the positivity of $%
c $ and $G_{ab}$, and vice versa. This is not completely surprising, because
we can expect the metric $f$ to be related to the normalization of the
two-point functions of scalar operators. From 5d supergravity, 
\begin{equation}
S= \int \sqrt{g}\left(-{\frac{R}{4}}+G_{ab}\partial\lambda^a\partial%
\lambda^b \right)  \label{norm}
\end{equation}
we can see by a simple scaling that, at least at the fixed points, where $%
ds^2=R^2[dy^2+\exp(2y)\sum_i dx_i^2]$ 
\begin{eqnarray}
\langle T(x) T(0)\rangle = {\frac{c}{|x|^8}}\,\,&\rightarrow& c\sim R^3\sim
(\Lambda )^{-3/2}  \nonumber \\
\langle \lambda_a (x) \lambda_b (0)\rangle = {\frac{f_{ab}}{|x|^{2\Delta}}}%
\,\,&\rightarrow& f_{ab}\sim R^3G_{ab}\sim cG_{ab}  \label{scaling}
\end{eqnarray}
The first equation reproduces the known result for $c$ \cite{hs}, the second
one confirms equation~(\ref{holometric}).

\textit{iii}) Using the arguments of \cite{a2} it is straightforward to show
that (\ref{holometric})\ defines a consistent \textit{scheme} choice, at
least when $c_{\mathrm{IR}}\neq 0$. We call this scheme the ``holographic
scheme'' and can be considered in the class of ``proper'' schemes of \cite
{a2}, in which the metric $f$ is set equal to a known, positive function:
the identity in \cite{a2}, $2cG^{ab}$ here. The choice $f_{ab}=\delta_{ab}$
defines the proper beta function $\beta_{\mathrm{P}}$ and relates the total $%
c$-flow to the area of the graph of the beta function. In the holographic
scheme, instead, we have (for $G^{ab}=\delta^{ab}$), $f^{ab}=2c\delta^{ab}$,
i.e. the total flow of $\ln c$ is (twice) the area of the graph of the
holographic beta function. In Fig. 1 the two beta functions are compared for
the model of \cite{gppz3}, involving the flow to the (confining) pure N=1
super-Yang Mills theory. The holographic beta function tends to a costant in
the IR, while the proper beta function better resembles an ordinary beta
function. When $c_{\mathrm{IR}}=0$, as in our last example, it is natural to
expect that the holographic scheme (\ref{holometric}) is still consistent,
because, although the holographic beta function tends to a constant, the
metric $f$ is zero in the ``null'' IR theory. This is what our explicit
computation of the $T_{\mu\nu}$ correlator shows.

\begin{figure}[tbp]
\centerline{\epsfig{figure=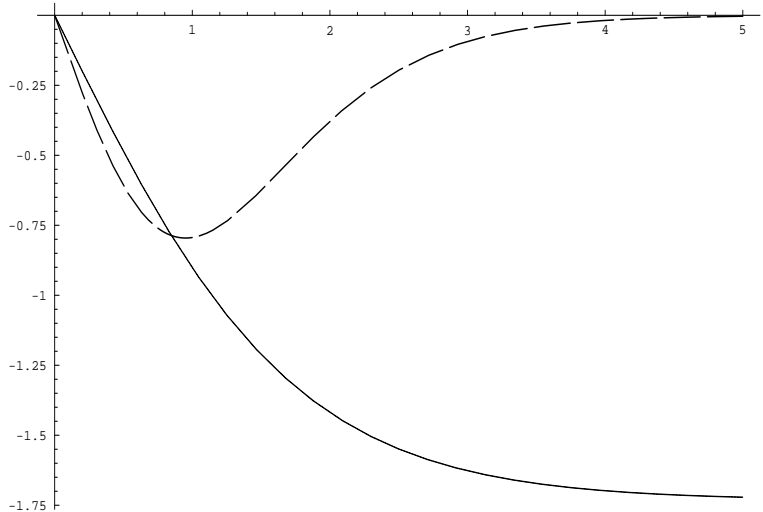,height=7 cm, width=10cm}}
\caption{Plot of the holographic (continuous line) and proper (dashed line)
beta functions versus the holographic coupling.}
\end{figure}

\textit{iv}) In the presence of many couplings the formula $\dot{c}=-2c\beta
_{a}\beta_bG^{ab}$ does not give all the beta functions separately. Yet, the
sum $\beta _{a}\beta_bG^{ab}$ is sufficient both to fix $h_{*}$ and to
identify the fixed points. In this sense we may call 
\begin{equation}
\beta _{\mathrm{H}}\equiv -\sqrt{\beta _{a}\beta_bG^{ab}}=-\sqrt{-\frac{\dot{%
c}}{2c}}
\end{equation}
the holographic beta function, so that $\dot{c}=-2c\beta _{\mathrm{H}}^{2}$.
The proper beta function is instead $\beta_{\mathrm{P}}=\beta_{\mathrm{H}}%
\sqrt{2c}$, so that $\dot c=-\beta_{\mathrm{P}}^2$.

\textit{v})\ With obvious changes, various formulas above apply for the $a$%
-function of \cite{a2} in the general case $c\neq a$. Indeed, the
relationship between the critical exponent $h_{*}$ and the $a$-function does
not require inputs from the AdS/CFT correspondence and holds purely in
quantum field theory. This generalization is straightforward and left to the
reader.

\textit{vi})\ In refs.~\cite{freed2}, explicit formulas for the two-point
function of minimally-coupled, massless scalars in the Coulomb branch of N=4
supersymmetric gauge theory are given. From those formulas, one can extract
a $c$-function using the same techniques described in this paper. As an
example we now briefly discuss the case of a a 4-dimensional distribution of
branes, giving rise to the two-point function described in eq.~(25) of ref.~%
\cite{freed2}. Following the same steps that led us to eq.~(\ref{int}) we
find the central function: 
\begin{equation}
c(x)={\frac{N^2}{4}}\sum_{n=2}^{\infty}(2n-1)\sqrt{n^2-n}|x|^3 K_1(2\sqrt{%
n^2-n}|x|).  \label{cgubs}
\end{equation}
This function is also positive and monotonic as shown in Figure 2. 
\begin{figure}[tbp]
\centerline{\epsfig{figure=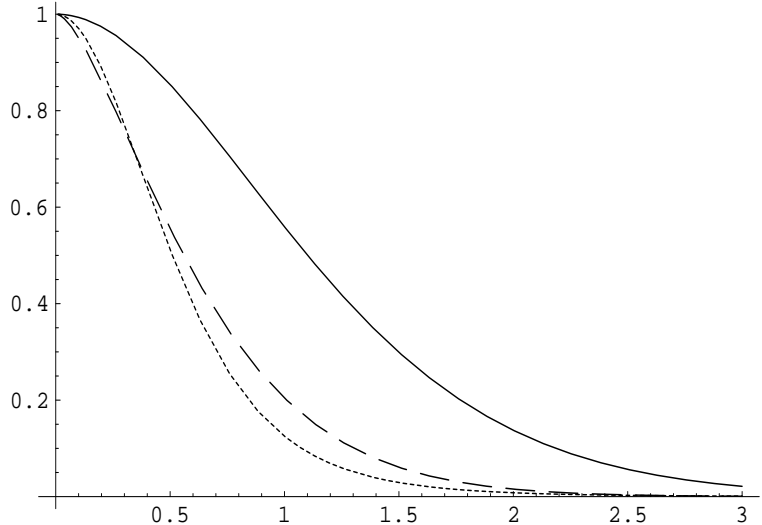,height=7 cm, width=10cm}}
\caption{Plot of $8c/N^2$ vs $x$, for different cases. In the case of the
flow to N=1 SYM, the canonical $c$ (eq.~(\ref{int})) is given by the dashed
line, and the holographic $c$ (eq.~(\ref{chol})) is given by the dotted
line. Finally, in the case of N=4 Coulomb branch, the canonical $c$-function
(eq.~(\ref{cgubs})) is given by the continuous line.}
\end{figure}

The holographic scheme is natural and simple. Other schemes and definitions
for $c$-functions are less natural from the point of view of the AdS/CFT
correspondence, but still have great interest in their own and give results
for $c=a$ theories that share many properties with 2d conformal field
theories. In particular, we considered the definition of a $c$-function from
the two-point function of the stress-energy tensor. We computed such a $c(x)$
for a particular supersymmetric flow. The fact that it is monotonic is a
highly non-trivial check of the AdS/CFT correspondence as well as of the
fact that supergravity solutions may be interpreted as quantum field theory
RG flows. Notice that the particular solution used in the computation is
singular in the IR (as it happens for all the cases where analytical
computations of two-point functions can be performed). Nevertheless, we
obtained a sensible result, which indicates that the basic physical
properties of such solutions are not completely spoiled by the IR
singularity.

We conclude by mentioning some possible extensions of this work that we find
particularly interesting. 1) To compute the two-point function for a
holographic flow between CFTs, as the one connecting the N=4 theory to an IR
N=1 CFT, discussed in ref.~\cite{freed1}. 2) To prove in full generality
that the canonical $c$-function is always monotonic, as it happens for the
holographic $c$-function. 3) To generalize formula (\ref{noncanonica}) to
the canonical $c$-function.

\vskip .2in \noindent \textbf{Acknowledgments}\vskip .1in \noindent We would
like to thank D. Z. Freedman, M. Petrini and A. Starinets for useful
discussions. L.G. and A. Z. are partially supported by INFN and MURST, and
by the European Commission TMR program ERBFMRX-CT96-0045, wherein they are
associated to the University of Torino. M.P. is supported in part by NSF
grant no. PHY-9722083.

\end{document}